\documentclass{elsart3}

\usepackage{graphicx}

\usepackage{amssymb}

\begin{document}

\begin{frontmatter}

\title{Magnetic order of FeMn alloy on the W(001) surface}


\author[]{Martin Ondr\'{a}\v{c}ek,}
\author[]{Josef Kudrnovsk\'{y}, and}
\author[]{Franti\v{s}ek M\'{a}ca\corauthref{cor1}}
\corauth[cor1]{Corresponding author.}
 \ead{maca@fzu.cz}

\address[]{Institute of Physics ASCR, Na Slovance 2, CZ--182 21 Prague 8, Czech Republic}


\begin{abstract}
We investigate  theoretically the ground state of the FeMn binary
alloy monolayer on the W(001) surface, the stability of different
magnetic configurations (ferro/antiferromagnetic, disordered local
moments, etc.) and estimate concentrations at which a transition
occurs between different magnetic orders. The tight-binding linear
muffin-tin orbital method combined with the coherent potential
approximation is used to treat the surface alloy appropriately. We
discuss the role of disorder on the phase transitions in surface
alloys composed from two different $3d$ transition metals.
\end{abstract}

\end{frontmatter}

\section{Introduction}
Magnetic phenomena at surfaces and interfaces have been studied
extensively in recent years. This interest in surface magnetism
has been stimulated by the needs of technology (magnetic storage
media, spintronics) as well as the development of tools for
research in this field, both experimental (e.g. the spin-polarized
scanning tunneling microscope) \cite{bode} and computational
(increasing speed and memory capacity of computers).

Reduced number of nearest neighbors at surfaces and lowering of
crystal symmetry due to the presence of the surface are known to
have a deep impact on magnetic properties. The magnetic moments of
atoms in the surface layer are often substantially enhanced as
compared to corresponding local magnetic moments inside the bulk.
 Elements that are known to be nonmagnetic in their bulk form can be
magnetic when placed on a surface.
An interesting example of how the chemical and geometrical
properties of a surface may influence the magnetism of an
adsorbate has been reported recently. Experiments indicated that
iron monolayers on the (001) tungsten surface had zero net
magnetization (see \cite{wulfhekel} and references therein),
in contrast, e.g., to iron monolayers on the W(110) surface that
are known to be ferromagnetic \cite{elmers}. The question whether
they were really non-magnetic or rather antiferromagnetic was
resolved by Kubetzka \cite{kubetzka} using the spin-polarized
scanning tunneling microscope. Here, antiferromagnetic  $c(2\times
2)$ Fe-monolayer on the W(001) was directly observed. The
subsequent theoretical study by  Ferriani
 \cite{ferriani} showed that iron and cobalt monolayers tend
to be antiferromagnetic on the W(001) surface while V, Cr and Mn
monolayers tend to the ferromagnetic order. It should be noted
that just an opposite trend is observed on the W(110) surface, on
which V, Cr, and Mn monolayers are $c(2\times 2)$ antiferromagnets
while Fe and Co order ferromagnetically. The same or similar
situation as on the W(110) surface occurs also on many other
substrates, e.g. on Cu, Ag and Pd (001) surfaces (see references
in \cite{ferriani}).

Our goal is to generalize the above study to the case of binary
two-dimensional alloys on the W(001) surface. The disorder in the
surface layer and the interaction between chemically different
surface atoms are expected to lead to new types of magnetic
phases. We have focused on the FeMn system in which alloy
constituents exhibits antiferromagnetic (Fe) and ferromagnetic
(Mn) order as pure overlayers. The frustration that results from
the competing interactions will play a key role in formation of
more complex magnetic structures.

\section{Method}
The electronic structure was described in the framework of the
tight-binding linear muffin-tin orbital (TB-LMTO) method with the
atomic sphere approximation (ASA) \cite{turek}. The spin-polarized
local density approximation (LDA) with the parametrization of
Vosko, Wilk, and Nusair \cite{vwn} for the exchange-correlation
potential was used. The spin-orbit interaction was neglected. The
surface Green function formalism was used to describe the
electronic structure of an overlayer on substrate with realistic
boundary conditions. The substitutional disorder was treated by
means of the coherent potential approximation (CPA), which
neglects local environment effects. In addition to the standard
ASA, a dipole surface barrier was included.

The geometry of the system is the body-centered cubic (bcc)
lattice of tungsten with the experimental lattice constant. The
self-consistent electronic structure calculations were performed
on an eight-layer slab: the slab consisted of four tungsten
layers, one layer of $3d$ transition-metal alloy, and three empty
layers that mimicked the vacuum above the surface. The slab was
then matched to a semi-infinite vacuum region on one side and to a
semi-infinite tungsten bulk on the other side. The layers of the
surface slab form an ideal continuation of the bulk structure and
possible layer relaxations are neglected.

We have tested the stability of various magnetic phases of
two-dimensional random overlayers on the W(001) substrate. In
addition to the ferromagnetic (FM) phase, we have also considered
the DLM phase (disordered local moments), the so-called APM phase
(antiparallel arrangement of moments), and the uncompensated DLM
(UDLM) phase.

The DLM phase is defined by randomly oriented local magnetic
moments. The distribution of directions is uniform in the whole
space. Such a structure can be efficiently studied using the
spin-polarized CPA: the infinite number of possible orientations
of local moments can be replaced by two equally large groups of
atoms with opposite orientations of local moments. The DLM phase
of a binary alloy X$_c$Y$_{1-c}$ can be formally described as a
four-component alloy
X$_{c/2}^{+}$X$_{c/2}^{-}$Y$_{(1-c)/2}^{+}$Y$_{(1-c)/2}^{-}$,
where the superscripts denote the orientations of respective local
moments. The DLM structure has a zero total magnetization
similarly as any antiferromagnetic structure. In contrast to an
ordered antiferromagnetic structure, however, the local moments in
the DLM phase are distributed on the lattice randomly. Energetic
preference of the DLM arrangement to the ferromagnetic one is an
indication that the antiferromagnetic arrangement would be favored
over ferromagnetism.

The APM arrangement in a binary alloy is defined as a magnetic
phase in which all local moments of one component are aligned in
one direction and all local moments of the other component are
aligned in the opposite direction. The APM phase can be described
as X$_{c}^{+}$Y$_{1-c}^{-}$. This arrangement resembles a
ferrimagnetic one in a similar way as the DLM resembles
antiferromagnetism. It should be noted that both the
antiferromagnet or ferrimagnet assume the existence of two
different sublattices while the DLM or APM arrangement exist on
one lattice.

A mixed magnetic state, which we denote as DLM(X)+FM(Y), can be
represented by a ternary compound
X$_{c/2}^{+}$X$_{c/2}^{-}$Y$_{1-c}^{+}$. The most general phase is
the uncompensated DLM state (UDLM). It can be approximated as a
four-component alloy X$^{+}$X$^{-}$Y$^{+}$Y$^{-}$ where all
occupation probabilities (concentrations), $c_X^+$, $c_X^-$,
$c_Y^+$, $c_Y^-$, are arbitrary with the obvious constraints
$c_X^+ + c_X^- = c$ and $c_Y^+ + c_Y^- = 1-c$. All previously
discussed phases can be viewed as special cases of the UDLM.

\section{Results and discussion}

The local magnetic moments of FM, DLM and APM phases of the
FeMn/W(001) alloy are shown in Fig.~1 and contrasted with behavior
of the FeV/W(001)
 system. Magnetic moments of Mn and also of Fe
are rigid, only weakly influenced by their environment.

\begin{figure}[htb]
\begin{center}
 \includegraphics[width=76mm,clip]{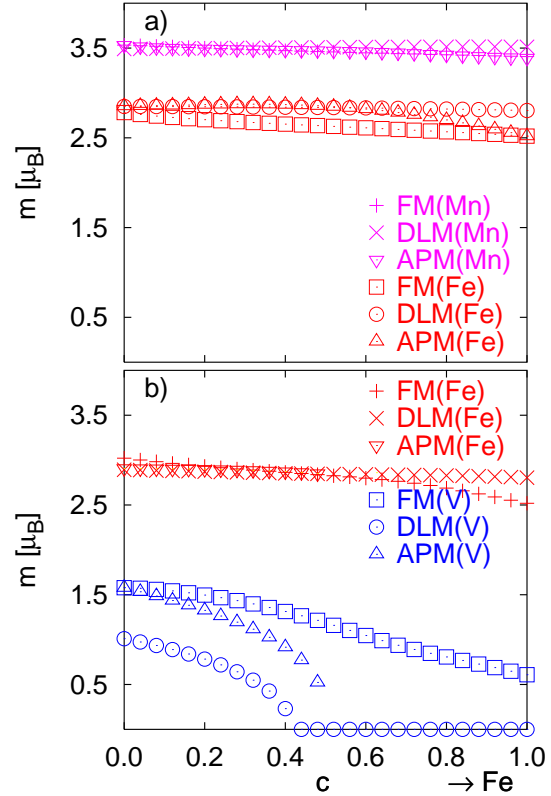}
\end{center}

\caption{a) Local magnetic moments at atoms of the FeMn surface
alloy as a function of iron concentration in various magnetic
states. b) The same for the FeV surface alloy.}
\end{figure}

Total energies for the FeMn/W(001) alloys in different magnetic
phases are presented in Fig.~2. To analyze it, let us consider
pairwise magnetic interactions only and suppose that the magnetic
interaction between two atoms is determined by the atoms
themselves independently of their environment. We can then
approximate the dependence of energy on magnetic configuration by
a simple quadratic form
\begin{eqnarray}\label{Eq1}
E - E^{DLM} & = & p (c_X^+ - c_X^-)^2 + q (c_Y^+ - c_Y^-)^2\nonumber \\
 &  &+ 2r (c_X^+ - c_X^-)(c_Y^+ - c_Y^-).
\end{eqnarray}
Specifically, we get:
\begin{eqnarray}
\label{FM} E^{FM} - E^{DLM} & = & p c_X^2 + q c_Y^2 + 2r c_X c_Y \\
\label{APM} E^{APM} - E^{DLM} & = & p c_X^2 + q c_Y^2 - 2r c_X c_Y \\
\label{DLMFM} E^{DLM+FM} - E^{DLM} & = & q c_Y^2.
\end{eqnarray}
The parameters $p$, $q$, and $r$ have been chosen to fit the
calculated total energies of the ferromagnetic Fe/W(001) surface,
the ferromagnetic Mn/W(001) surface, and the
Fe$_{0.5}$Mn$_{0.5}$/W(001) surface in the APM phase.

\begin{figure}[htb]
\begin{center}
\begin{tabular}{c}
\includegraphics[width=78mm,clip]{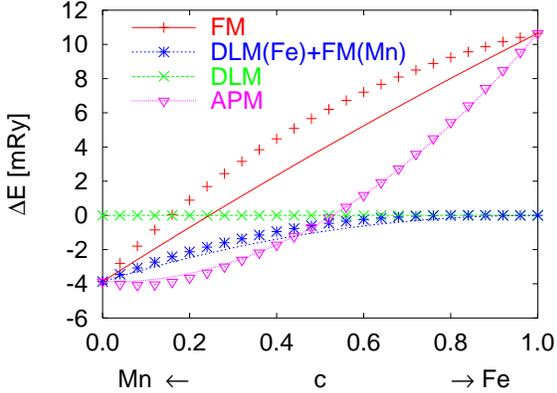}
\end{tabular}
\end{center}
\caption{Total energies of various magnetic phases of the FeMn
surface alloy with respect to the reference DLM state as a
function of iron concentration. Symbols denote ab-initio
calculated values, lines show values obtained from the model
described in the text.}
\end{figure}

As can be seen from Fig.~2, the DLM(Fe)+FM(Mn) phase has the
lowest energy in the iron-rich FeMn overlayers. The DLM(Fe)+FM(Mn)
phase allows both elements to retain their preferred magnetic
state. In manganese-rich overlayers, the APM phase has lower total
energy as compared to the DLM+FM phase. In the APM phase, the Mn
local moments keep their parallel (ferromagnetic) arrangement
while the Fe local moments are oriented antiparallel with respect
to the Mn ones. We can trace three tendencies that drive the
magnetic behavior of the FeMn/W(001) system: (i) local moments of
Fe atoms tend to be oriented antiparallel to each other, (ii)
local moments of Mn atoms tend to align parallel to each other,
and (iii) the local moments of Fe tend to be oriented antiparallel
to Mn moments.

The parameters in (\ref{Eq1}) express the strength of pair
magnetic interactions between atoms in the overlayer: $p$
describes the interactions between Fe atoms, $q$ between Mn atoms,
and $r$ between Fe and Mn atoms. The qualitative agreement between
this model and the self-consistently calculated results (cf.
Fig.~2) justifies our discussion of the results in terms of the
three driving tendencies that correspond to these three
parameters. Our preliminary studies indicate that this model is
applicable to other $3d$ transition metal alloy overlayers with
rigid magnetic moments, but not to, e.g., vanadium alloys where
more complex behavior has been found.

\begin{table}[htb]
\begin{center}
\begin{tabular}[b]{|c|c|c|c|c|c|}
\hline \bf composition & $\mathbf{c_{X}^{+} / c_X}$ &
$\mathbf{c_{Y}^{+} / c_Y}$ &
$\mathbf{\bar{m}_X [\mu_{B}]}$ & $\mathbf{\bar{m}_Y [\mu_{B}]}$ &
{\bf state} \\
\hline
Fe$_{0.25}$Mn$_{0.75}$ & 1.00 & 0.00 & 2.87 & 3.49 & APM \\
Fe$_{0.50}$Mn$_{0.50}$ & 0.75 & 0.00 & 2.85 & 3.49 & UDLM \\
Fe$_{0.75}$Mn$_{0.25}$ & 0.60 & 0.00 & 2.83 & 3.50 & UDLM \\
\hline
\end{tabular}
\end{center}\vglue 2mm

\caption{\label{tabu} UDLM states with the lowest total energy for
FeMn surface binary alloys. First column gives the chemical
composition. Then the magnetic arrangement with the lowest energy
is described; $\bar{m}_X [\mu_{B}]$ and $\bar{m}_Y [\mu_{B}]$ are
average local moments of corresponding elements. The abbreviation
in the last column characterizes the magnetic state qualitatively.
}
\end{table}

We have also investigated the UDLM states to find the state with
the lowest total energy. Good sampling of possible UDLM
configurations requires a large number of calculations. We have
therefore performed calculations for only three different
compositions of FeMn (see Table~\ref{tabu}). It should be noted
that for all concentrations Mn moments prefer ferromagnetic
alignment (${c_{Y}^{+} / c_Y}$~=~0).

For some surface alloy compositions, there are UDLM states that
have lower total energy than the special magnetic arrangements
discussed so far. We have verified that the formula (\ref{Eq1})
describes fully self-consistent calculations for FeMn alloys with
a good accuracy even in the UDLM phase under the assumption that
parameters $p$, $q$, and $r$ are determined from calculated total
energies of the DLM, FM, APM, and DLM+FM phases.  The existence of
UDLM phases is a consequence of the interplay between the three
basic trends discussed above.

For pure Fe and Mn monolayer is the first interlayer distance
reduced by 13.7\% and 4.7\%, respectively \cite{ferriani}. To
judge the impact of interlayer relaxation on our results, we have
reduced distances between FeMn alloy addlayer and the first
tungsten layer by 10\%. This relaxation has no remarkable
influence on magnetic behavior of FeMn alloy film. Nevertheless,
the surface relaxation suppress magnetic moments of iron and
manganese atoms slightly because of $d$-band broadening due to
stronger bonding with substrate.
\section{Conclusions}
We have investigated magnetic ground states of two-dimensional
binary FeMn transition-metal  alloys on W(001). For pure
monolayers we have found that Fe overlayer order
antiferromagnetically while Mn atoms prefer the ferromagnetic
alignment. In alloy monolayers, Fe and Mn tend to orient their
local magnetic moments antiparallel to each other.

Incompatible tendencies of magnetic moment alignments among Fe and
Mn atoms lead to an uncompensated DLM phase, which becomes the
calculated ground state for some alloy compositions. This
indicates tendency to a non-collinear ground state. To reach a
more detailed understanding of the alloyed monolayers on W(001),
further SP-STM measurements as well as theoretical calculations
are needed.\\

\noindent {\bf Acknowledgments}

{This work has been done within the project AVOZ1-010-0520 of the
ASCR. We acknowledge the support from the Grant Agency of the
Czech Republic, Contract No. 202/04/583, and the COST P19-OC150
project.}

\end{document}